\documentclass[aip,jap,twocolumn]{revtex4-1}

\usepackage{graphicx}

\begin{document}

\title{Universal $R$-$C$ crossover in current-voltage
characteristics for unshunted array of overdamped $Nb-AlO_x-Nb$
Josephson junctions}

\author{S. Sergeenkov}

\affiliation{Materials and Devices Group, Department of Physics
and Physical Engineering, Universidade Federal de S\~ao Carlos,
S\~ao Carlos, SP, 13565-905 Brazil}


\author{V.A.G. Rivera}

\affiliation{Materials and Devices Group, Department of Physics
and Physical Engineering, Universidade Federal de S\~ao Carlos,
S\~ao Carlos, SP, 13565-905 Brazil}

\author{E. Marega}

\affiliation{Instituto de F\'{\i}sica de S\~ao Carlos, USP, Caixa
Postal 369,  S\~ao Carlos, SP, 13560-970 Brazil}

\author{F.M. Araujo-Moreira}

\affiliation{Materials and Devices Group, Department of Physics
and Physical Engineering, Universidade Federal de S\~ao Carlos,
S\~ao Carlos, SP, 13565-905 Brazil}


\revised{23 March 2010}

\begin{abstract}
We report on some unusual behavior of the measured current-voltage
characteristics (CVC) in artificially prepared two-dimensional
unshunted array  of overdamped $Nb-AlO_x-Nb$ Josephson junctions.
The obtained nonlinear CVC are found to exhibit a pronounced (and
practically temperature independent) crossover at some current
$I_{cr}=\left(\frac{1}{2\beta_C}-1\right)I_C$ from a resistance
$R$ dominated state with $V_R=R\sqrt{I^2-I_C^2}$ below $I_{cr}$ to
a capacitance $C$ dominated state with
$V_C=\sqrt{\frac{\hbar}{4eC}} \sqrt{I-I_C}$ above $I_{cr}$. The
origin of the observed behavior is discussed within a
single-plaquette approximation assuming the conventional RSJ model
with a finite capacitance and the Ambegaokar-Baratoff relation for
the critical current of the single junction.
\end{abstract}



\maketitle

Among many different properties which can be studied using highly
ordered two-dimensional arrays of Josephson junctions probably one
of the most interesting (and important for their potential
applications) is their transport behavior, reflecting
manifestation of numerous dissipation mechanisms in the arrays
(see, e.g.,~\cite{1,2,3} and further references therein) via
nonlinear current-voltage characteristics (CVC) of the general
form $V\propto [I-I_C(T)]^{a(T)}$ with a power exponent $a$
ranging from $a<1$ to $a>1$ depending on the particular mechanism
and collateral effects (such as finite size of the single junction
and/or the array, thermal fluctuations, quasi-particle
contributions, etc~\cite{4,5,6,7,8,9}). At the same time, recall
that only sufficiently overdamped Josephson junctions (with
nonhysteretic CVC) and their arrays can be effectively used in
rapid single flux quantum (RSFQ) logic circuits and programmable
Josephson voltage standards (see, e.g.,~\cite{a,b,c,d} and further
references therein). In this paper we report our results on CVC
for SIS type array of strongly overdamped $Nb-AlO_{x}-Nb$
junctions at different temperatures. We observed quite a
pronounced crossover at some current
$I_{cr}=\left(\frac{1}{2\beta_C}-1\right)I_C$ between a resistance
$R$ dominated state (below $I_{cr}$) and a capacitance $C$
dominated state (above $I_{cr}$) which could be utilized as a
versatile $R$-$C$ switch within Josephson electronics. High
quality ordered SIS type unshunted array of overdamped
$Nb-AlO_{x}-Nb$ junctions has been prepared by using a standard
photolithography and sputtering technique~\cite{1}.  It is formed
by loops of niobium islands linked through $100\times 150$ tunnel
junctions. The unit cell of the array has square geometry with
lattice spacing $a\simeq 46\mu m$ and a single junction area of
$5\times 5\mu m^{2}$. The critical current for the junctions
forming the arrays is $I_C(T)=150 \mu A$ at $T=1.7K$. Given the
values of the junction quasi-particle resistance $R=10\Omega$ and
capacitance $C=1.2fF$, the circuit frequency and dissipation
measuring Stewart-McCumber parameter are estimated to be $\omega
_{RC}=1/CR\simeq 10^{14}Hz$ and $\beta _{C}(T)=\frac{2\pi
CR^{2}I_{C}(T)}{\Phi _{0}}\simeq 0.05$ at $T=1.7K$, respectively.
The parameters of the array are as follows, $I_{CA}(1.7K)=1.2mA$
and $R_A=25\Omega$. The measurements were made using homemade
experimental technique with a high-precision
nanovoltmeter~\cite{8}.
\begin{figure}
\centerline{\includegraphics[width=7.5cm]{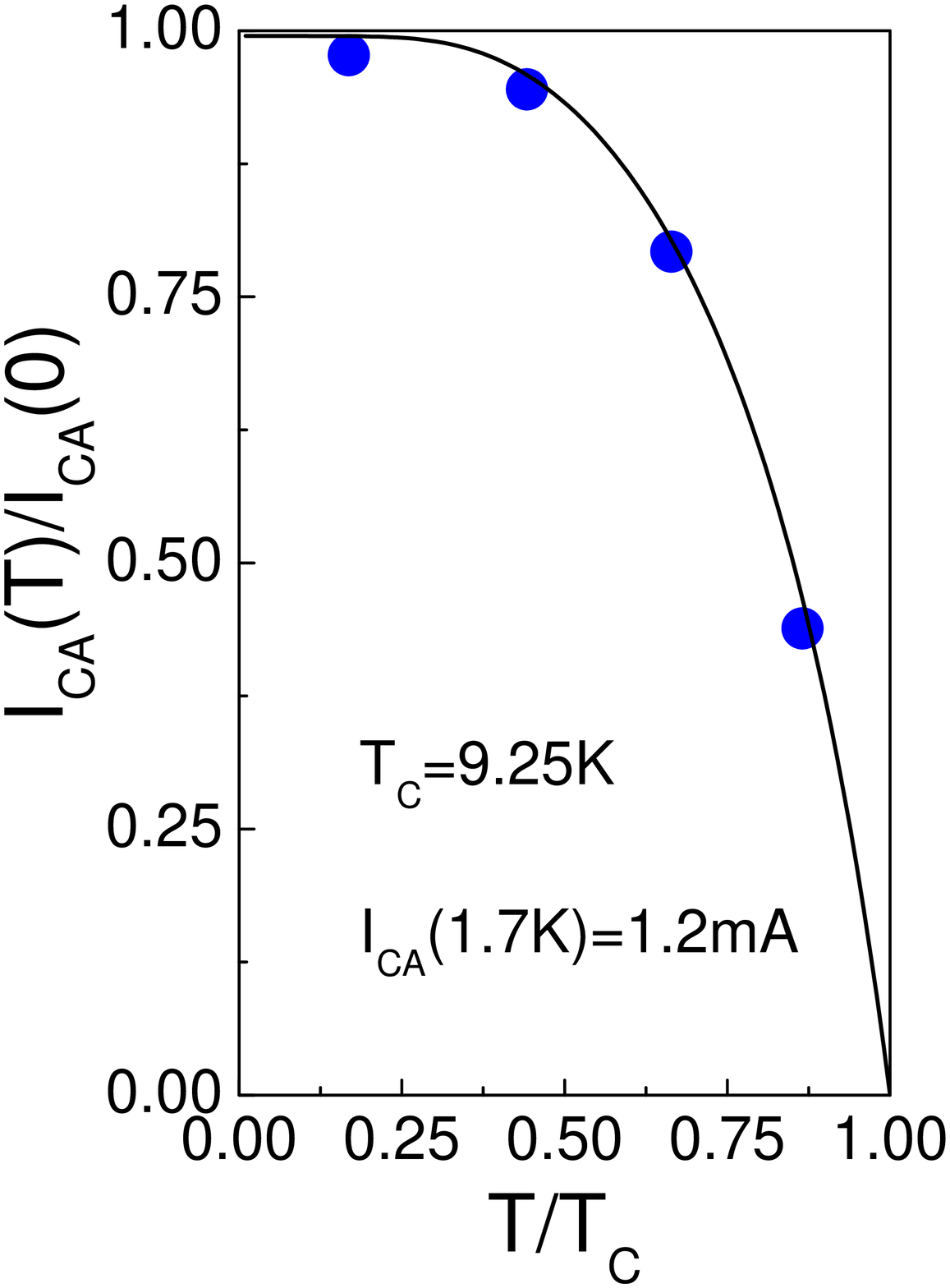}}\vspace{0.25cm}
\caption{\label{fig:1} (Color online) The temperature dependence
of the critical current for the array of overdamped
$Nb-AlO_{x}-Nb$ Josephson junctions. The solid line is the best
fit using the single-junction Ambegaokar-Baratoff relation.}
\end{figure}
The temperature dependence of the normalized critical current of
the array $I_{CA}(T)/I_{CA}(0)$ is shown in Fig.1. Observe that it
is very well fitted (solid line) to the Ambegaokar-Baratoff
relation~\cite{10} for the critical current of a {\it single
junction} $I_{C}(T)=I_{C}(0)\left[ \frac{\Delta (T)}{\Delta
(0)}\right] \tanh \left[ \frac{\Delta (T)}{2k_{B}T}\right]$ where
$\Delta (T)=\Delta (0)\tanh
\left(2.2\sqrt{\frac{T_{C}-T}{T}}\right)$ is the analytical
approximation~\cite{11} of the BCS gap parameter (valid for all
temperatures) with $\Delta (0)=1.76k_BT_C\simeq 1.5meV$ and
$I_C(0)=\pi \Delta (0)/2eR \simeq 150 \mu A$. This remarkable
experimental fact suggests a rather strong coordinated response
from all the junctions forming the array and allows us to
substantially simplify the analysis of the obtained results by
considering the properties of a single junction (or plaquette, see
below). For this purpose,  the initial CVC data for array were
rescaled by directly introducing the critical current of the
single junction $I_C$. Some typical results of the normalized
rescaled CVC taken at different temperatures are shown in Fig.2.
The solid lines through the data points are the best fits
according to the following expressions:
$V_R=V_0\sqrt{(I^2-I_C^2)/I_0^2}$ for $I_C<I<I_{cr}$ and
$V_C=V_0\sqrt{(I-I_C)/I_0}$ for $I>I_{cr}$ with $V_0=30mV$ and
$I_0=3mA$. It is interesting to point out that, according to
Fig.2, the crossover shows almost a universal behavior taking
place around $[I-I_C(T)]/I_0\simeq 0.4$ for {\it all}
temperatures.
\begin{figure*}
\begin{center}
\includegraphics[width=7.5cm]{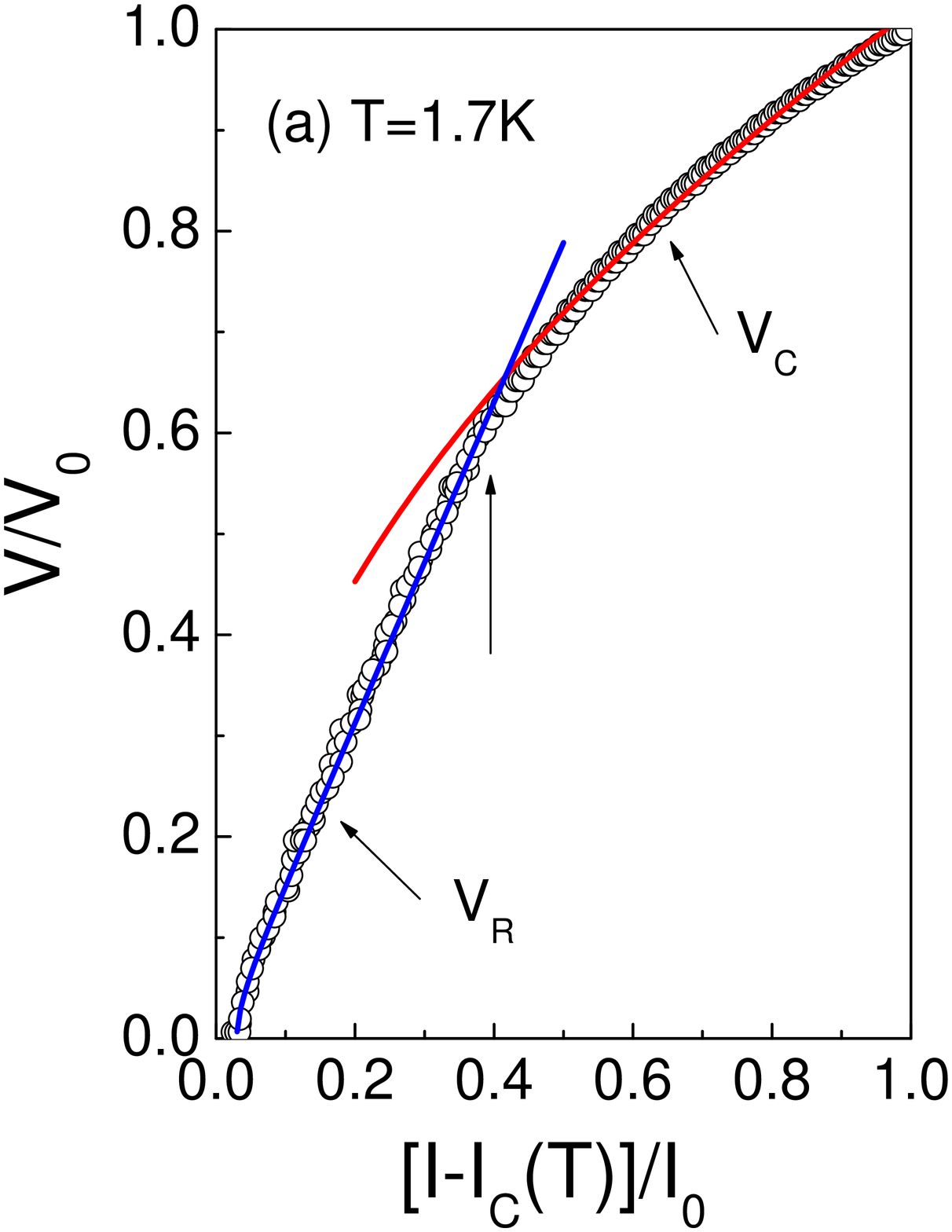}
\includegraphics[width=7.5cm]{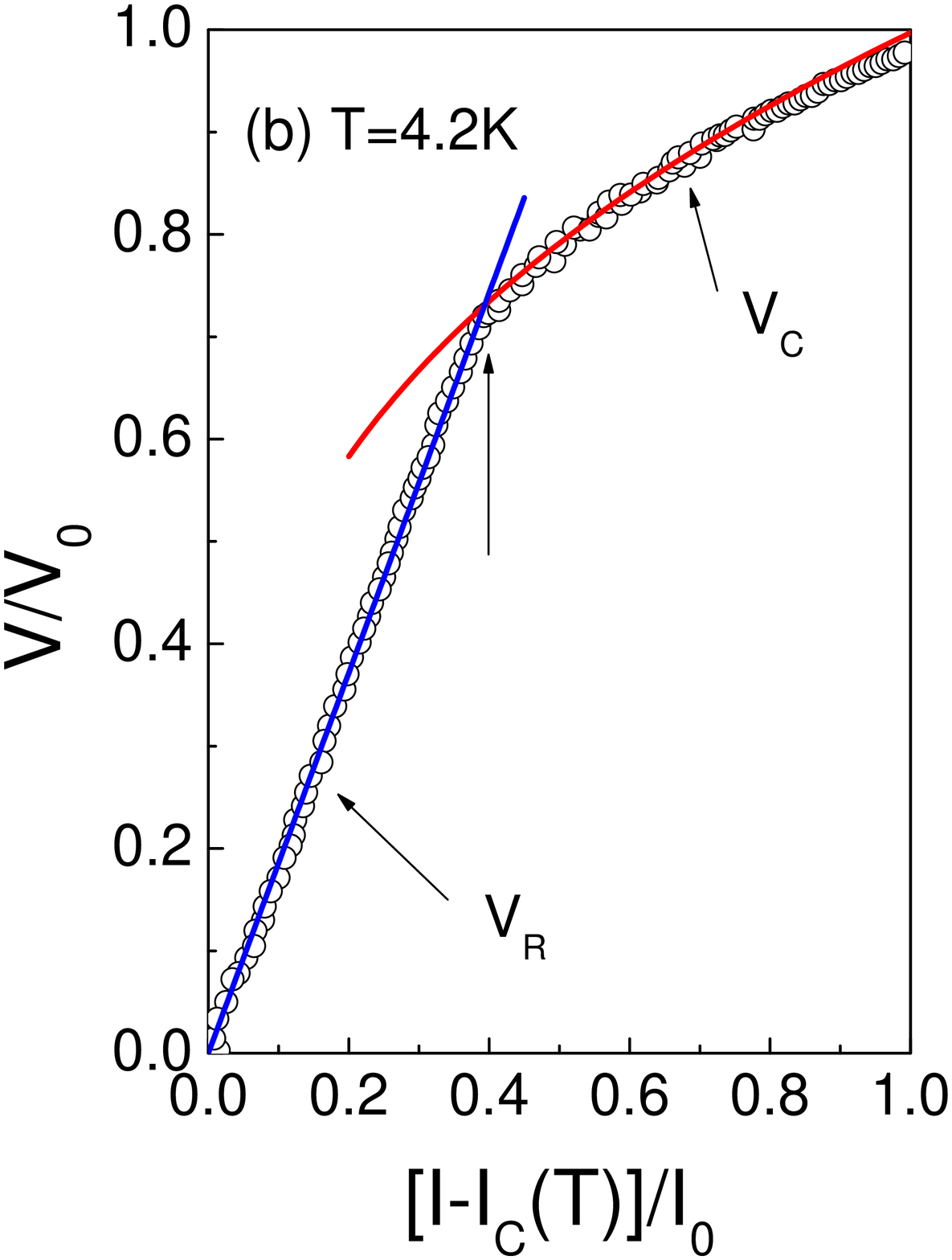}
\includegraphics[width=7.5cm]{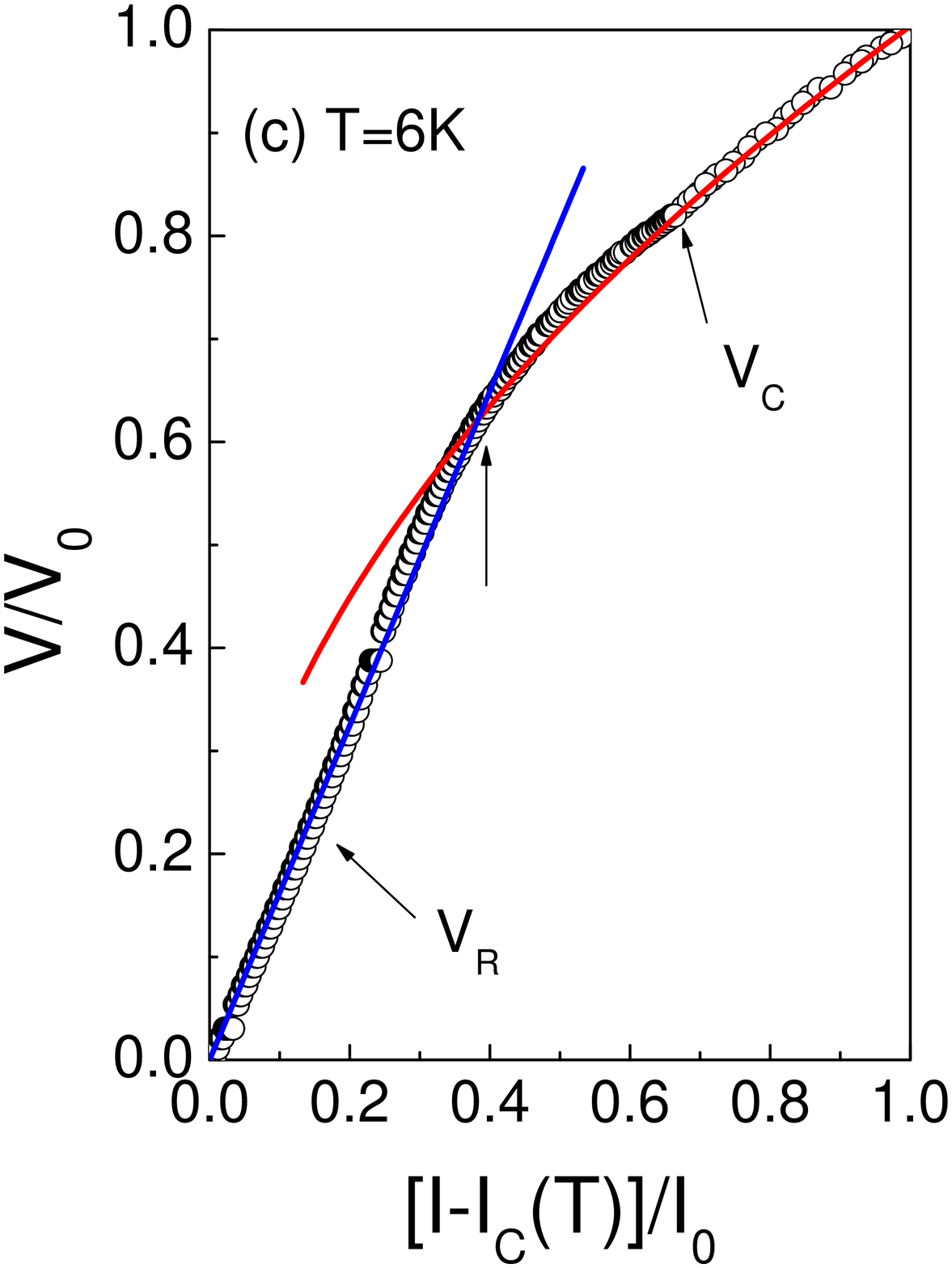} \includegraphics[width=7.5cm]{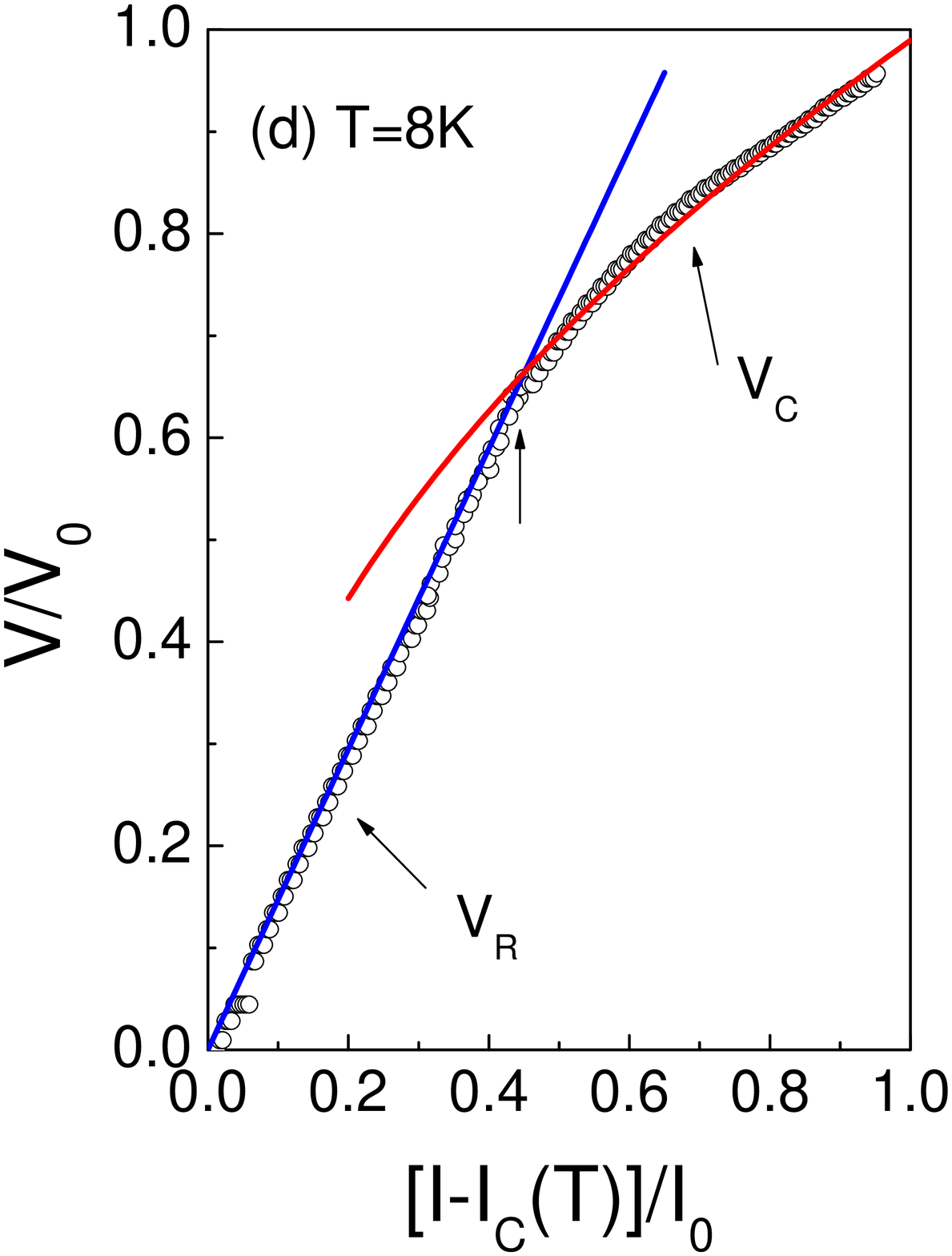}
\caption{\label{fig:2} (Color online) The normalized rescaled
current-voltage characteristics for unshunted array of overdamped
$Nb-AlO_{x}-Nb$ Josephson junctions taken at various temperatures
along with the best fits (solid lines) using expressions for $V_R$
and $V_C$ (see text). Here, $I_0=3mA$ and $V_0=30mV$.}
\end{center}
\end{figure*}
To understand the observed behavior of the CVC in our array, in
principle one would need to analyze in detail the dynamics of the
array. However, as we have previously
reported~\cite{12,13,14,15,16}, because of the well-defined
periodic structure of our array, it is reasonable to expect that
our experimental results can be quite satisfactory explained by
analyzing the dynamics of a single unit cell of the array. In our
calculations, the unit cell is a plaquette containing four
identical Josephson junctions. By analogy with the resistively
shunted junction (RSJ) model~\cite{9}, the total current in the
plaquette reads~\cite{12} $I=I_C(T)\sin \phi _i(t)+\frac{\Phi
_0}{2\pi R}\frac{d\phi _i}{dt}+ \frac{C\Phi _0}{2\pi}\frac{d^2\phi
_i}{dt^2}$. Here $\phi _i(t)$ is the gauge-invariant
superconducting phase difference across the $i$th junction, and
$\Phi _0$ is the magnetic flux quantum. For any particular
solution $\phi _i(t)$ of this equation at $I\neq 0$, the resulting
CVC of the RSJ model is given by the time average of the voltage
($\tau=2\pi /\omega$ is properly defined period)
$V(I)=\frac{\hbar}{2e\tau}\int_0^{\tau}dt\left(\frac{d\phi
_i}{dt}\right)$. For the resistance dominated situation (when the
capacitance related effects can be totally neglected, that is when
$\omega_{RC}\tau \gg 1$), the RSJ model has a well-known
solution~\cite{9} $\phi _i(t)=2\tan^{-1}\left[\frac{\hbar
\omega}{2eRI}\tan\left(\frac{\omega t
}{2}\right)-\frac{I}{I_C}\right]$ with
$\omega=\frac{2eR}{\hbar}\sqrt{I^2-I_C^2}$, which brings about
$V_R=\frac{\hbar}{2e\tau}\int_0^{\tau}dt\left(\frac{d\phi
_i}{dt}\right)=R\sqrt{I^2-I_C^2}$ for $R$-dominated CVC. As we can
see, this dependence exactly corresponds to the fitting expression
for the observed CVC below $I_{cr}$ assuming the Ohmic relation
$V_0=RI_0$. Let us turn now to the opposite situation and consider
the capacitance dominated regime when the resistance related
effects can be totally neglected (that is when $\omega_{RC}\tau
\ll 1$). In this case, the first integral of the RSJ model reads
$\frac{d\phi _i}{dt}=\sqrt{2}\omega
_p\sqrt{\frac{C_1}{I_C}+\frac{I}{I_C}\phi_i+\cos \phi_i}$ where
$\omega _p=\sqrt{2eI_C/\hbar C}$ is the plasmon frequency, and
$C_1$ is the integration constant. Unfortunately, for $I\neq 0$,
this equation can not be solved exactly. Hence, let us consider
its approximate solution assuming that
$\phi_i(t)=\frac{\pi}{2}+\theta(t)$ with $\theta (t) \ll
\frac{\pi}{2}$. By fixing the arbitrary constant as
$C_1=-\frac{\pi}{2}I$, within this approximation, we obtain
$\theta(t)=\Omega^2t^2$ with $\Omega =\omega
_p\sqrt{(I-I_C)/2I_C}$, which in turn results in the following
explicit form of $C$-dominated CVC,
$V_C=\frac{\hbar}{2e\tau}\int_0^{\tau}dt\left(\frac{d\theta}{dt}\right)=
\sqrt{\frac{\hbar}{4eC}}\sqrt{I-I_C}$. As we can see, this
dependence exactly corresponds to the fitting expression for the
observed CVC above $I_{cr}$ assuming $V_0=\omega
_p\sqrt{2eI_0/\hbar C}=\hbar \omega _{RC}/2e$ and
$I_0=V_0/R=I_C/\beta_C$ for the normalization parameters.
Furthermore, given the experimental values for $R$ and $C$, we
obtain $V_0\simeq 30mV$ and $I_0\simeq 3mA$. In turn, from the
obvious identity $V_R(I_{cr})=V_C(I_{cr})$, we readily obtain
$I_{cr}=(\frac{1}{2\beta_C}-1)I_C$ for the crossover current.
Finally, a comment is in order regarding the employed here
simplified model of $2D$ array. It should be noted that we
completely ignored inductance (geometry) related effects which
seem to be of less importance for the interpretation of the
observed crossover than dissipation induced factors (resistance
and capacitance). However, for more adequate description of the
flux dynamics in {\it truly} $2D$ systems, these effects should be
taken into account. Indeed, as accurate numerical simulations have
revealed~\cite{17a,17b}, both self-inductance and mutual
inductance effects will have a significant impact on the array's
dynamic properties through creation of rather strong self-induced
magnetic fields.

We are indebted to R.S. Newrock, P. Barbara and C.J. Lobb for
helpful discussions and for providing SIS samples. This work has
been financially supported by the Brazilian agencies CNPq, CAPES
and FAPESP.

\nocite{*}
\bibliography{aipsamp}

\begin{thebibliography}{99}

\bibitem{1} R.S. Newrock, C.J. Lobb, U. Geigenmuller, and M. Octavio,  Solid
State Physics {\bf 54}, 263 (2000).

\bibitem{2} P. Martinoli, C. Leeman, J. Low Temp. Phys. {\bf 118}, 699
(2000).

\bibitem{3} R. Fazio and  H. van der Zant, Phys. Rep. {\bf 355}, 235
(2001).

\bibitem{4} J.M. Repaci, C. Kwon, Qi Li, Xiuguang Jiang, T. Venkatessan, R.E. Glover III, C.J. Lobb,
and R.S. Newrock, Phys. Rev. B {\bf 54}, R9674 (1996).

\bibitem{5}  S.W. Pierson, M. Friesen, S.M. Ammirata, J.C. Hunnicutt, and L.A. Gorham,
Phys. Rev. B {\bf 60}, 1309 (1999).

\bibitem{6} K. Medvedyeva, B.J. Kim, and P. Minnhagen, Phys. Rev. B {\bf
62}, 14531 (2000).

\bibitem{7}  J. Holzer, R.S. Newrock, C.J. Lobb, T. Aouaroun, and S.T. Herbert, Phys. Rev. B {\bf 63}, 184508
(2001).

\bibitem{8}  V.A.G. Rivera,  S. Sergeenkov, E. Marega, and
F.M. Araujo-Moreira, Phys. Lett. A {\bf 374}, 376 (2009).

\bibitem{9} A. Barone and  G. Paterno, \textit{Physics and Applications of the
Josephson Effect} (A Wiley-Interscience Publisher, New York,
1982).

\bibitem{a} J. Niemeyer, in {\it Handbook of Applied Superconductivity}, edited by
B. Seeber (IOP, Bristol, 1998), p. 1813.

\bibitem{b} T. Van Duzer, L. Zheng, X. Meng, C. Loyo, S.R. Whiteley, L.Yu, N. Newman,
J.M. Rowell, and N. Yoshikawa, Physica C {\bf 372}-{\bf 376}, 1
(2002).

\bibitem{c} J. Kohlmann, R. Behr, and T. Funck, Meas. Sci. Technol. {\bf 14}, 1216
(2003).

\bibitem{d} K. Senapati and Z.H. Barber, Appl.
Phys. Lett. {\bf 94}, 173511 (2009).


\bibitem{10} V. Ambegaokar  and  A. Baratoff, Phys. Rev. Lett. {\bf 10}, 486
(1963).

\bibitem{11}  S. Sergeenkov, JETP Lett. {\bf 76}, 170 (2002).

\bibitem{12} F.M. Araujo-Moreira, P. Barbara, A.B. Cawthorne, and C.J. Lobb,
 Phys. Rev. Lett. {\bf 78}, 4625 (1997).

\bibitem{13} P. Barbara, F.M. Araujo-Moreira, A.B. Cawthorne, and C.J. Lobb,
Phys. Rev. B {\bf 60}, 7489 (1999).

\bibitem{14} S. Sergeenkov and F.M. Araujo-Moreira, JETP Lett. {\bf 80}, 580
(2004).

\bibitem{15} F.M. Araujo-Moreira, W. Maluf, and S. Sergeenkov,  Eur. Phys. J.
B {\bf 44}, 25 (2005).

\bibitem{16}  F.M. Araujo-Moreira and S. Sergeenkov, Supercond. Science and
Technol. {\bf 21}, 045002 (2008).

\bibitem{17a} J.R. Phillips, H.S.J. van der Zant, and T.P. Orlando,
Phys. Rev. B {\bf 50}, 9380 (1994).

\bibitem{17b} J.R. Phillips, H.S.J. van der Zant, J. White, and T.P. Orlando,
Phys. Rev. B {\bf 50}, 9387 (1994).

\end{thebibliography}

\end{document}